\documentclass[twocolumn]{aastex631}

\graphicspath{{./}{figures/}}
\usepackage{amsmath}
\usepackage{threeparttable}
\usepackage{graphicx}        % For resizing the table
\usepackage{times}
\usepackage{CJK}
\usepackage{tablefootnote}

\newcommand\sref[1]{\hyperref[#1]{\S~\ref*{#1}}}
\newcommand\fref[1]{\hyperref[#1]{Fig.~\ref*{#1}}}
\newcommand\Eqref[1]{Eq.~(\hyperref[#1]{\ref*{#1}})}
\newcommand\eeqref[1]{Eq.~\hyperref[#1]{\ref*{#1}}}
\newcommand\tref[1]{\hyperref[#1]{Table~\ref*{#1}}}
\newcommand\aref[1]{\hyperref[#1]{Appendix~\ref*{#1}}}

\shorttitle{From Dispersion to Metallicity}
\shortauthors{Roy et al.}

\begin{document}
\begin{CJK}{UTF8}{mj}
%\UseRawInputEncoding
%\linenumbers
%\linenumbers
\title{From Dispersion to Metallicity: Probing the Milky Way Circumgalactic Medium with Fast Radio Bursts}

\correspondingauthor{Manami Roy}
\email{roy.516@osu.edu}
\author[0000-0001-9567-8807]{Manami Roy}
\affiliation{Center for Cosmology and Astro-Particle Physics, The Ohio State University, 191 W. Woodruff Avenue, Columbus, OH 43210, USA}
\affiliation{Department of Astronomy, The Ohio State University, 140 W. 18th Ave., Columbus, OH 43210, USA}
\affiliation{Department of Physics, The Ohio State University, 191 West Woodruff Ave., Columbus, OH 43210, USA}

\author[0000-0002-9069-7061]{Sanskriti Das}
\altaffiliation{Hubble Fellow}
\affil{Kavli Institute for Particle Astrophysics and Cosmology, Stanford University, 452 Lomita Mall, Stanford, CA\,94305, USA}
\email{snskriti@stanford.edu; dassanskriti@gmail.com}

\author[0000-0002-4822-3559]{Smita Mathur}
\affiliation{Center for Cosmology and Astro-Particle Physics, The Ohio State University, 191 W. Woodruff Avenue, Columbus, OH 43210, USA}
\affiliation{Department of Astronomy, The Ohio State University, 140 W. 18th Ave., Columbus, OH 43210, USA}
\author{Anjali Gupta}
\affil{Columbus State Community College, 550 E Spring Street, Columbus, OH 43210, USA}
\author[0000-0001-6291-5239]{Yair Krongold}
\affil{Instituto de Astronomia, Universidad Nacional Autonoma de Mexico, 04510 Mexico City, Mexico}

\begin{abstract} \label{abstract}
Circumgalactic Medium (CGM) plays a critical role in galactic evolution, serving as the primary reservoir for accretion and outflows. In this work, we present a novel statistical analysis to derive CGM metallicity from the dispersion measure  (DM) of 25 Fast Radio Burst (FRB) sightlines. By combining metallicity-dependent oxygen absorption DM estimates with metallicity-independent FRB DMs, we directly constrain the metallicity of the hot phase of the Milky Way (MW) CGM. Considering different diffuse baryon fractions ($f_{\text{d}} = 0.0, 0.6, 0.8$) in the Intergalactic Medium (IGM) and intervening halos, the median metallicities of CGM are $Z_{\text{CGM}} = 0.12_{-0.04}^{+0.09} \, Z_{\odot}$, $0.26_{-0.12}^{+0.18} \, Z_{\odot}$, and $0.39_{-0.20}^{+0.26} \, Z_{\odot}$, respectively, by the bootstrapping method. Super-solar metallicity is seen in multiple sightlines for $f_d=0.8$, which is implausible for MW CGM metallicities. Even under conservative lower-bound estimates ($f_{\text{d}} = 0.0$), the inferred CGM metallicity significantly exceeds expectations for IGM accretion ($Z_{\text{IGM}} \lesssim 0.1 \, Z_{\odot}$), providing strong empirical evidence that a large fraction of the metal-enriched galactic feedback material is efficiently retained within the Milky Way CGM. Furthermore, we observe significant sightline-to-sightline variation in inferred metallicities, pointing toward a highly anisotropic multi-phase CGM shaped by complex interactions between feedback outflows and cosmic accretion.
\end{abstract}

% \PN{PN: will rewrite afterward}
%\keywords{methods: numerical --- galaxies: evolution --- Galaxy: halo}

\section{Introduction}
\label{S:intro}
%https://ui.adsabs.harvard.edu/abs/2025ApJS..280....6C/abstract
% to do -- log normal and list the other FRB
The Circumgalactic Medium (CGM) is a diffuse gaseous halo surrounding the galactic disk that acts as a vital interface for the cosmic baryon cycle \citep{Tumlinson2017}. Continuously replenished by metal-rich galactic feedback from supernova-driven outflows and accretion from the Intergalactic Medium (IGM), the CGM governs galaxies' long-term star formation history and chemical evolution \citep{Tumlinson2017,Fau2023}. This diffuse gaseous halo is multiphase in nature, spanning from cold ($T \lesssim 10^4\text{K}$) to hot, volume-filling components ($T \gtrsim 10^6~\text{K}$).

While soft X-ray spectroscopy (e.g., O~\textsc{vii} and O~\textsc{viii} absorption line profiles) effectively probes the high-temperature $>10^6~\text{K}$ phase \citep{Mathur2022}, the X-ray spectra lack accessible neutral hydrogen transitions along the same pencil-beam sightlines to directly determine total hydrogen column densities. Consequently, converting observed metal-ion column densities into total gas masses requires assuming an arbitrary halo metallicity (typically $0.1\text{--}0.3 \, Z_{\odot}$) \citep{Toft2002,Sommer2006,Tumlinson2017,Fau2023}, leaving the true metallicity of the hot, volume-filling CGM fundamentally unconstrained.

%Mathur, S. (2022). Probing the Circumgalactic Medium with X-Ray Absorption Lines. In: Bambi, C., Santangelo, A. (eds) Handbook of X-ray and Gamma-ray Astrophysics. Springer, Singapore. https://doi.org/10.1007/978-981-16-4544-0_112-1

In this work, we present a novel empirical framework to directly measure the metallicity of the hot CGM phase by exploiting the rapidly growing population of localized Fast Radio Bursts (FRBs) \citep{Bhardwaj2021,Bhardwaj2021a,Bhardwaj2024,Zhang2024,Catalouge_FRB2025}. Because local FRBs propagate through minimal cosmic path lengths, their total dispersion measure ($\text{DM} = \int n_{\text{e}} \, \mathrm{d}l$)) is dominated by the Milky Way (MW) and host galaxy halos rather than the diffuse IGM. 

There have been several studies that tried to model the contribution of DM from the MW CGM. Accurately constraining the DM of the MW CGM ($\text{DM}_{\text{MW, CGM}}$) is pivotal for both CGM and cosmological studies. For the CGM community, it breaks the mass-metallicity degeneracy in soft X-ray absorption analyses, directly quantifying the diffuse halo baryon reservoir and feedback retention efficiency. For the FRB community, precisely isolating this Galactic foreground is essential to calibrate the cosmological Macquart ($\text{DM}_{\text{cosmic}}$--$z$) relation and probe intrinsic host-galaxy environments.\cite{Yamasaki2020} constructed a combined disk plus spherical halo density profile calibrated against Galactic X-ray emission measures ($\text{EM} = \int n_{\text{e}} n_{\text{p}} \, \mathrm{d}l$; See also \cite{Keating2020}). While this approach significantly improves upon disk-only models \cite[e.g.,][]{Cordes2002, Yao2017} at high Galactic latitudes ($|b| > 20^{\circ}$), converting EM to DM introduces substantial systematic uncertainties. Crucially, EM traces the square of the electron density, causing it to miss the diffuse, low-density volume of the hot CGM that can only be reliably mapped in absorption. Consequently, EM-inferred DM contributions strongly depend on poorly constrained parameters, including the halo's spatial extent, virial mass, and total baryon fraction \citep{Gupta2012, Boylan2013,Gupta2017}. 

Alternatively, absorption-line analyses offer a direct path to MW DM calculation by measuring column densities ($N_{\text{X}} = \int n_{\text{X}} \, \mathrm{d}l$). \cite{Prochaska2019} utilized $\text{O}~\textsc{vii}$ $\text{K}\alpha$ absorption columns from \cite{Fang2015} to constrain the hot Galactic halo's DM. 
However, because current instrument resolutions (such as XMM-Newton RGS) cannot resolve line velocity widths, observed line equivalent widths are susceptible to saturation effects, introducing additional errors in the inferred column densities. Instead, \cite{Das2021} combined the equivalent widths of the OVII K-$\alpha$ and OVII K-$\beta$ lines to constrain the saturation-corrected column density and the velocity width. Instead of assuming $f_{\rm OVII}$=1 everywhere, they used the OVIII column density to estimate $f_{\rm OVII}$. For this paper, we use the DM measurement of the MW CGM using absorption by \cite{Das2021}.

However, the DM estimates for the MW CGM derived from X-ray absorption column measurements inherently depend on the assumed gas metallicity, as they rely on converting specific metal-ion column densities across multi-phase ionization states into total electron column density. Conversely, the DM of the MW CGM measured directly from FRBs captures the total integrated free-electron column density along the line of sight and depends poorly on chemical composition. By taking the ratio of the metallicity-dependent absorption DM estimate to the metallicity-independent FRB DM, we propose a novel technique to calculate the metallicity of the MW CGM. Assuming the IGM contribution is negligible for these local FRBs, this method can directly constrain the metallicity ($\text{(O/H)}$) of the hot phase of the CGM.
We describe our methods in Section \ref{sec:methods}, following the demonstration of the result of this paper in Section \ref{sec:result}. Section \ref{sec:conclusion} provides a short discussion and conclusion.
%Briefly discuss other DM estimates of MW CGM in the literature (I cited them in Das+2021), and why ours is better :) It shouldn't look like we haven't considered other methods, although many people exclude ours :/ 
\section{Methods}
\label{sec:methods}

\begin{figure}
\includegraphics[width=\linewidth]{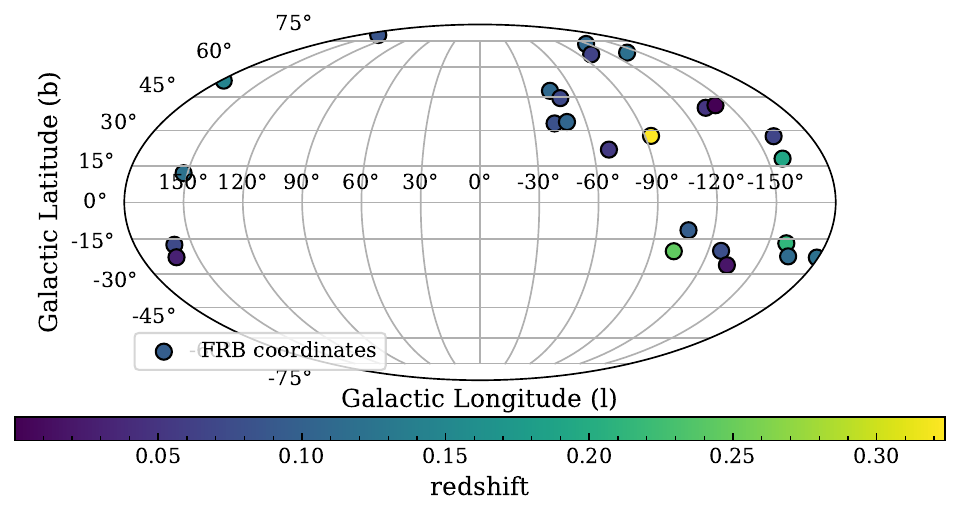}
\caption{The coordinates and redshift of the Sample of local-universe FRBs used for this work.}
\label{fig:frb_l_b}
\end{figure}

In this section, we outline the step-by-step methodology used to infer the metallicity of the MW CGM. Our sample includes the local FRBs from \cite{Bhardwaj2021, Bhardwaj2021a, Bhardwaj2024, Zhang2024, Catalouge_FRB2025}. We have noted their name, latitude, longitude, redshift, and total observed DM($\text{DM}_{\text{obs}}$) in the first five columns of Table \ref{tab:local_FRB_list_3fds}. Figure \ref{fig:frb_l_b} shows their Galactic positions, color-coded with their redshifts. 

\subsection{{Halo DM Contribution from X-ray Observations}}\label{subsec:xray_dm}

To account for the multi-phase nature of the MW's CGM, \citet{Das2021} derived empirical constraints on the Galactic dispersion measure ($\text{DM}$) by combining DM of cold, cool, warm, and hot phases probed by 21-cm, ultraviolet (UV), extreme ultraviolet (EUV), and X-ray observations, respectively: 
\begin{equation}
    \rm DM_{\rm Gal} = \rm DM_{\rm cold} + \rm DM_{\rm cool} + \rm DM_{\rm warm} + \rm DM_{\rm hot} 
\end{equation}
where ``cold'' refers to $\lesssim10^4$ K gas which is predominantly neutral, ``cool'' is $10^{4-5}$K mildly ionized gas (probed by low-ions like SiII, SiIII), ``warm'' is for $\approx 10^{5-5.5}$ K gas probed by Li-like ions (primarily OVI), and hot refers to  $\geqslant 10^6$K gas probed by H- and He-like ions, e.g., OVII and OVIII.    

Their findings demonstrated that the total halo $\text{DM}$ is heavily dominated by the hot gas phase ($T \sim 10^6\text{ K}$), directly probed via X-ray absorption lines of OVII (and OVIII, if available) as measured by \citep{Gupta2012,Fang2015,Nicastro2016,Gupta2017,Gatuzz2018}. The other phases contribute $\sim15\%$ to the total DM value on average.

An observed ionic column density $N_{\text{metal}}$ converts to an equivalent electron dispersion measure via:
\begin{equation}
\text{DM}_{\text{metal}} = \int n_e \, dl \approx \left(\frac{n_e}{n_\text{H}}\right) \frac{N_{\text{metal}}}{f_{\text{metal}} \, A_{\text{X},\odot} \, (Z_{ref}/Z_\odot)}
\label{eq:dm_conversion}
\end{equation}
where $f_{\text{metal}}$ is the ion fraction, $A_{\text{X},\odot}$ is the solar elemental abundance relative to hydrogen, and $Z_{ref}/Z_\odot$ is the gas metallicity normalized to solar. Following \citet{Das2021}, we adopt the hot-halo component ($\text{DM}_{\text{hot}}$) parameterized as:
\begin{equation}
\begin{aligned}
   \text{DM}_{\text{hot}} = 83.7\text{ cm}^{-3}\text{ pc} \left(\frac{N_{\text{O\,VII}}}{10^{16.5}\text{ cm}^{-2}}\right)  \left(\frac{f_{\text{O\,VII}}}{1}\right)^{-1}  \\ \times \left(\frac{A_{\text{O},\odot}}{4.9\times10^{-4}}\right)^{-1} \left(\frac{Z_{[O/H]}}{0.3 Z_\odot}\right)^{-1} 
\end{aligned}
\end{equation}

%***I'm commenting out the next line; it should go in the Introduction.***
%X-ray absorption spectra lack hydrogen lines to yield direct abundance measurements, 
In the above equation, $Z = 0.3 Z_\odot$ is used, but this is an assumption that we will test.  
%physically motivated by warm-gas condensation via hot-phase cooling. 
Rather than assuming $f_{\text{O\,VII}} = 1$, \citet{Das2021} derived $f_{\text{O\,VII}}$ directly from the temperature (or its upper limit) $N_{\text{O\,VIII}}/N_{\text{O\,VII}}$ column density ratio under collisional ionization equilibrium (CIE), if both OVII and OVIII (or its upper limit) are detected along a sightline. Where $\text{O\,VIII}$ measurements are unavailable due to low signal-to-noise or instrumental features, $f_{\text{O\,VII}} = 1$ is adopted, as the peak CIE fraction ($\approx 0.9$) falls well within observational uncertainties.

%To overcome the curve-of-growth degeneracy between $N_{\text{O\,VII}}$ and the Doppler parameter $b$ caused by $\text{O\,VII}\ \text{K-}\alpha$ line saturation \citep{Draine2011}, $N_{\text{O\,VII}}$ is constrained by finding the unique intersection of $N(b)$ contours derived from both the $\text{O\,VII}\ \text{K-}\alpha$ and $\text{K-}\beta$ lines. Conversely, $\text{O\,VIII}$ lines remain optically thin, allowing $N_{\text{O\,VIII}}$ to scale linearly with equivalent width.

\subsection{Halo DM Contribution from FRB Observations}
\label{subsec:frb_dm}

The total observed DM from a localized FRB can be decomposed into four distinct line-of-sight components:
\begin{equation}
\text{DM}_{\text{obs}} = \text{DM}_{\text{MW, ISM}} + \text{DM}_{\text{MW, CGM}} + \text{DM}_{\text{cosmic}} + \frac{\text{DM}_{\text{host}}}{1+z},
\end{equation}
where $\text{DM}_{\text{MW, ISM}}$ and $\text{DM}_{\text{MW, CGM}}$ represent the interstellar and CGM of the MW, $\text{DM}_{\text{cosmic}}$ is the contribution from the IGM and intervening halos, and $\text{DM}_{\text{host}}$ encompasses the combined contribution from the progenitor environment, host galaxy ISM, and host CGM. 

 We restrict our sample to localized FRBs at $z < 0.3$. Although the intergalactic contribution is subdominant to the host term for many of
these sources, it is not negligible. Adopting a flat $\Lambda\text{CDM}$ universe ($H_0 = 70~\mathrm{km\,s^{-1}\,Mpc^{-1}}$, $\Omega_{\rm m} = 0.31$, $\Omega_{\rm b} = 0.049$,$\Omega_{\rm \Lambda} = 0.685$), the average cosmological dispersion measure contribution $\langle\text{DM}_{\text{cosmic}}\rangle$ to redshift $z_{\text{FRB}}$ is given by \cite{Mac2020}:
\begin{equation}
\langle\text{DM}_{\text{cosmic}}\rangle = \int_{0}^{z_{\text{FRB}}} \frac{c \, \bar{n}_{\text{e}}(z) \, \mathrm{d}z}{H_{0} (1+z)^{2} \sqrt{\Omega_{\text{m}} (1+z)^{3} + \Omega_{\Lambda}}}
\end{equation}
where the mean electron density $\bar{n}_{\text{e}}(z)$ incorporates the fraction of cosmic baryons in diffuse ionized gas, $f_{\text{d}}(z)$:
\begin{equation}
\bar{n}_{\text{e}}(z) = f_{\text{d}}(z) \, \rho_{\text{b}}(z) \, m_{\text{p}}^{-1} \left(1 - \frac{Y_{\text{He}}}{2}\right)
\end{equation}
Here, $m_{\text{p}}$ is the proton mass, $Y_{\text{He}} = 0.25$ is the helium mass fraction (assumed doubly ionized), and $\rho_{\text{b}}(z) = \Omega_{\text{b}} \rho_{\text{c,0}} (1+z)^{3}$ is the cosmic baryon mass density. The parameter $f_{\text{d}}(z)$ accounts for the fraction of baryons locked in diffuse ionized phases  outside the host and the MW (mainly in IGM and intervening halos):
\begin{equation}
f_{\text{d}}(z) = 1 - \left( f_{\text{stars}}(z) + f_{\text{remnants}}(z) + f_{\text{ISM}}(z) \right)
\end{equation}
We treat
$f_{\text{d}}(z)$ as constant and bracket its value with $f_{\rm d} = 0.6$ and $0.8$,
together with $f_{\rm d} = 0$ as a limiting case in which all
extragalactic DM is attributed to the host.
%***then f_d is the baryon fraction in the diffuse phase. This is contrary to the statement above.***
For the Galactic ISM component ($\text{DM}_{\text{MW, ISM}}$), we employ standard electron density model of NE2025 \citep{NE2025}. 

To account for uncertainties ($\text{DM}_{\text{host}}$) and the intrinsic scatter of the FRB host galaxy contribution, we take the model for the rest-frame host dispersion measure ($\text{DM}'_{\text{host}}$) using a log-normal probability density function (PDF)  \citep{Mac2020,James2022}. Let $x$ be a continuous random variable that follows a normal (Gaussian) distribution with mean $\mu_{\text{host}}$ and standard deviation $\sigma_{\text{host}}$:
\begin{equation}
x \sim \mathcal{N}(\mu_{\text{host}}, \sigma_{\text{host}}^2).
\end{equation}
The corresponding probability density function $f_X(x)$ is given by:
\begin{equation}
f_X(x) = \frac{1}{\sigma_{\text{host}}\sqrt{2\pi}} \exp \left[ -\frac{(x - \mu_{\text{host}})^2}{2\sigma_{\text{host}}^2} \right].
\label{eq:gaussian_pdf}
\end{equation}

We define the physical host dispersion measure in the rest frame via the exponential variable transformation $x = \ln(\text{DM}'_{\text{host}})$, or equivalently $\text{DM}'_{\text{host}} = e^x$. 

Therefore, the standard log-normal probability density function for $\text{DM}'_{\text{host}}$:
\begin{equation}
\begin{aligned}
p(\text{DM}'_{\text{host}}) = &\frac{1}{\text{DM}'_{\text{host}}}   \frac{1}{\sigma_{\text{host}}\sqrt{2\pi}} \\ &
\times \exp \left[ -\frac{(\ln \text{DM}'_{\text{host}} - \mu_{\text{host}})^2}{2\sigma^2_{\text{host}}} \right].
\end{aligned}
\label{eq:lognormal_pdf_final}
\end{equation}

To propagate the statistical uncertainty of the host contribution ($\text{DM}_{\text{host}}$) into our inferred CGM metallicity ($Z$), we execute a Monte Carlo (MC) sampling procedure over $N$ realizations for each FRB in our sample. For each iteration $i$, a rest-frame host contribution $\text{DM}'_{\text{host}, i}$ is drawn from the log-normal distribution defined in Equation~\eqref{eq:lognormal_pdf_final}, parameterized by $e^{\mu_{\text{host}}} = 100.0$ and $\sigma_{\text{host}} = 1.0$, considering favorable values used by \cite{Mac2020, James2022, Connor2025}.  The observer-frame host contribution is calculated via:

\begin{equation}
\text{DM}_{\text{host}, i} = \frac{\text{DM}'_{\text{host}, i}}{1+z} 
\end{equation}
which yields an implied Milky Way CGM dispersion measure of:
\begin{equation}
\text{DM}_{\text{MW, CGM}, i} = \text{DM}_{\text{obs}} - \text{DM}_{\text{MW, ISM}}  - \text{DM}_{\text{cosmic}} - \text{DM}_{\text{host}, i}.
\end{equation}

The physical column densities must be strictly positive; therefore, we apply a physical boundary condition requiring $\text{DM}_{\text{MW, CGM}, i} > 0$. Realizations yielding unphysical values ($\text{DM}_{\text{MW, CGM}, i} \le 0$) are rejected. {For each sightline, Monte Carlo sampling was performed until exactly $N = 10^5$ valid physical realizations ($\text{DM}_{\text{MW, CGM}} > 0$) were obtained.} For the valid realizations, the corresponding metallicity $Z_i$ is evaluated using below Equation~\eqref{eq:z_final_ratio}. 

%We report the final CGM metallicity as the median of the accepted MC distribution, with $16\text{th}$ and $84\text{th}$ percentiles defining the lower and upper $68\%$ confidence intervals ($\pm 1\sigma$ equivalent errors):

    %Assumes $\text{DM}_{\text{host}}$ follows a log-normal distribution with a mean of $\mu = 2.08$ and a standard deviation of $\sigma = 0.5$ (in $\log_{10}$ space), representative of empirical FRB population statistics \citep{Mac2020,James2022}.

%\begin{figure}
%\includegraphics[width=0.33\textwidth]{plots/mass_1e7.pdf}
%\includegraphics[width=\linewidth]{DM_host.pdf} \label{fig:frb_dist}
%\caption{kk}
%\end{figure}

\begin{figure}
\includegraphics[width=\linewidth]{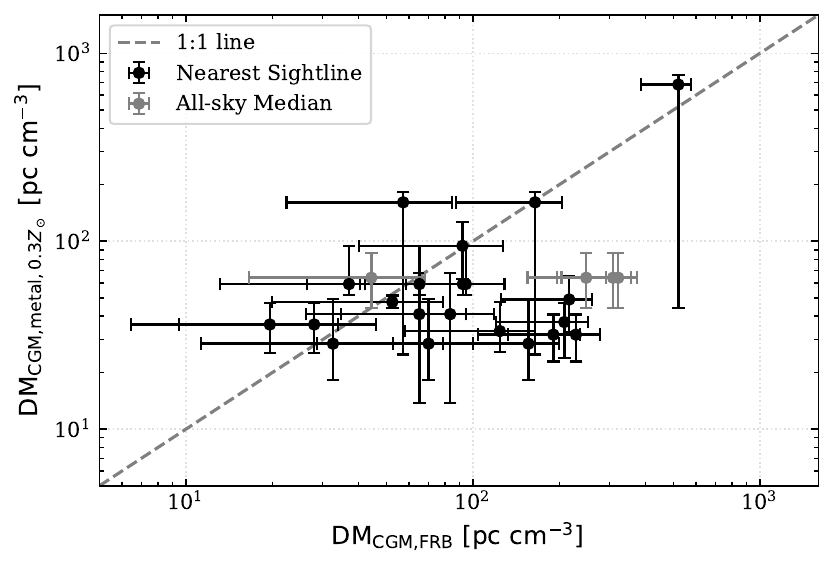} 
\caption{DM of CGM derived from metal column densities (assuming hot gas metallicity to be 0.3 solar metallicity) vs. derived from FRB observations, considering IGM contribution to be 0. The gray points denote the all-sky median where the nearest absorption sightlines to FRBs are not found.}
\label{fig:frb_dist}
\end{figure}

\section{Result}
\label{sec:result}
To infer the absolute metallicity of the circumgalactic medium ($Z$), we constructed the ratio between the expected hot-phase dispersion measure evaluated at a reference metallicity ($\text{DM}_{\text{X-ray, } Z_{\text{ref}}}$) and the empirical halo dispersion measure derived from FRBs ($\text{DM}_{\text{MW, CGM}}$). To estimate the $\text{DM}_{\text{X-ray,} Z_{\text{ref}}}$ and its associated uncertainty along each FRB line of sight, we used method of \cite{Das2021} who evaluated angular separations to all absorption sightlines using three complementary matching schemes: (i) adopting the measurement from the single nearest sightline within an angular separation threshold of $\theta_{\text{th}} = 0.5^{\circ}$, (ii) computing the mean and median of all sightlines within an angular threshold of $\theta_{\text{th}} = 0.5^{\circ}$, and (iii) taking the mean and median of sightlines located within an offset speration tolerance of $\Delta\theta_{\text{tol}} = 0.1^{\circ}$ relative to the closest sightline. 

First, we define the reference X-ray dispersion measure assuming a benchmark metallicity $Z_{\text{ref}}$. Since the true physical dispersion measure probed by FRBs ($\text{DM}_{\text{MW, CGM}}$) is basically independent of metallicity, the actual gas metallicity scales inversely with the electron column required to produce the observed ionic absorption (from Equation \ref{eq:dm_conversion}):
\begin{equation}
\frac{Z_{\text{MW,CGM}}}{Z_{\text{ref}}} = \frac{\text{DM}_{\text{CGM, metal, } Z_{\text{ref}}}}{\text{DM}_{\text{CGM, FRB}}}.
\label{eq:z_ratio}
\end{equation}

In Figure~\ref{fig:frb_dist}, we compare the sightline-to-sightline variation of MW CGM DM estimates derived from FRB observations ($\text{DM}_{\text{FRB}}$) against those inferred from metal absorption lines ($\text{DM}_{\text{metal}}$) along corresponding nearest sightlines. The uncertainty in $\text{DM}_{\text{metal}}$ ($\sigma_{\text{metal}}$) is primarily governed by errors in the measured absorption column densities, whereas the uncertainty in $\text{DM}_{\text{FRB}}$ ($\sigma_{\text{FRB}}$) is dominated by variations in the host galaxy contribution ($\text{DM}_{\text{host}}$). For FRBs lacking nearby absorption sightlines, we adopt the all-sky median absorption DM value, indicated by gray data points. The solid diagonal line represents a one-to-one correspondence where the two estimates align under an assumed reference metallicity of $Z_{\text{ref}} = 0.3 \, Z_{\odot}$. Data points lying systematically above (or below) this reference line indicate that the true local CGM metallicity $Z_{\text{MW, CGM}}$ is higher (or lower) than the reference value $Z_{\text{ref}}$.

%*** Discuss what the figure shows: 1:1 correspondance but with sightline to sightline variations***

\begin{figure*}
\includegraphics[width=\linewidth]{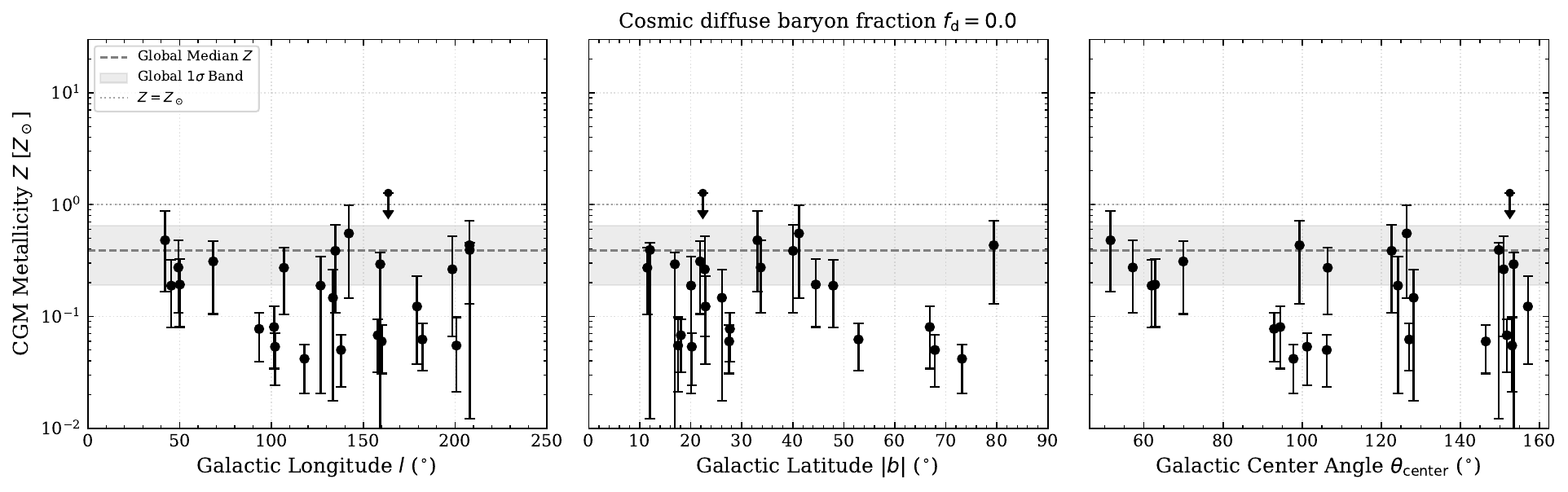}
\includegraphics[width=\linewidth]{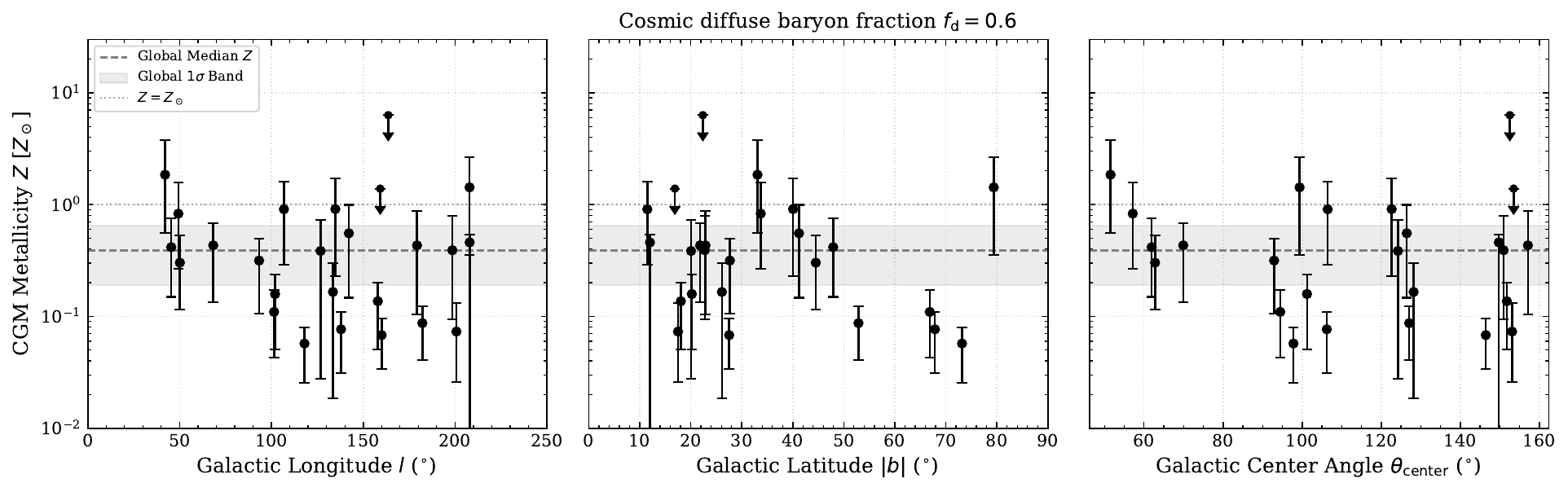}
\includegraphics[width=\linewidth]{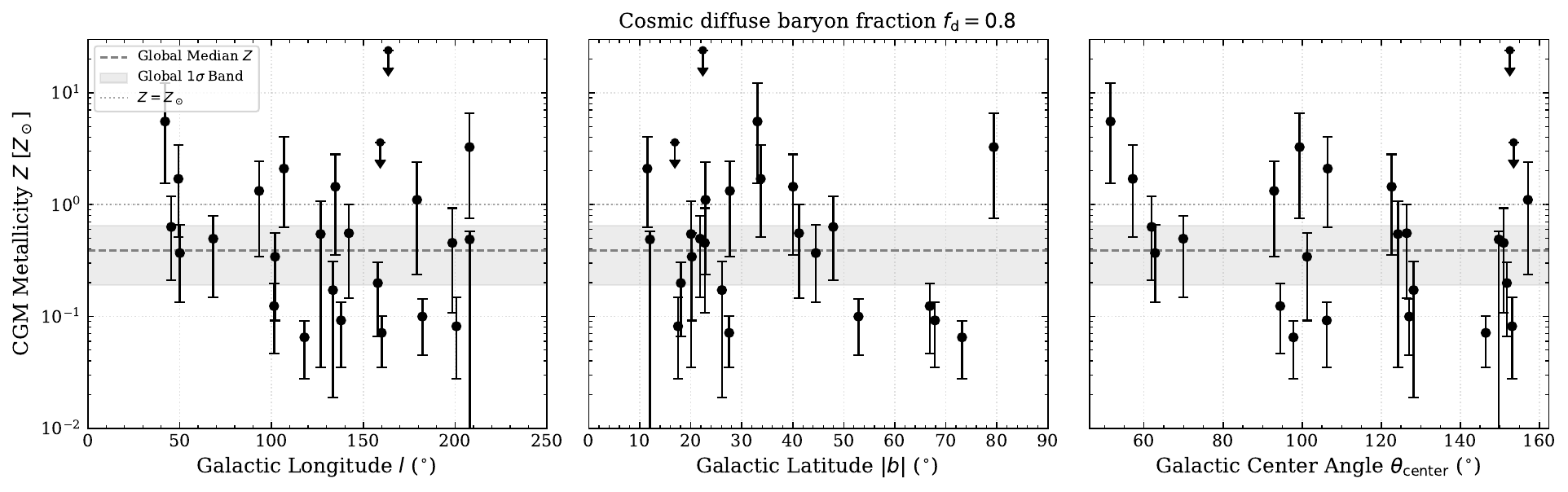}
\caption{CGM metallicity variation with longitude (left), latitude (middle), and distance from the Galactic Center along different sightlines. The three rows denote contributions from three different cosmic diffuse baryon fractions. Dashed line and shaded region in each plot denotes global median and 1$\sigma$ error respectively.} 
\label{fig:metals}
\end{figure*}

Expanding the FRB halo contribution into its constituent line-of-sight components yields the final operational expression:
\begin{equation}
Z_{\text{MW,CGM,i}} = Z_{\text{ref}} \left[ \frac{\text{DM}_{\text{metal, } Z_{\text{ref}}}}{\text{DM}_{\text{obs}} - \text{DM}_{\text{MW, ISM}} - \dfrac{\text{DM}_{\text{host}}}{1+z}} \right].
\label{eq:z_final_ratio}
\end{equation}
Figure~\ref{fig:metals} illustrates the derived CGM metallicities and their associated asymmetric uncertainties across the analyzed sightlines. The left, middle, and right panels display $Z_{\text{CGM}}$ as a function of Galactic longitude ($l$), Galactic latitude ($b$), and angular distance from the Galactic Center ($\theta_{\mathrm{center}}$), respectively. While we detect no clear global or radial trend in halo metallicity, significant sightline-to-sightline variations are evident, pointing toward an anisotropic and spatially inhomogeneous metal distribution within the MW CGM. We also list our derived metallicity values along with their errors in the last three columns of Table \ref{tab:local_FRB_list_3fds}.

The three rows highlight the impact of varying the diffuse IGM baryon fraction ($f_{\text{d}}$). Under the assumption of zero IGM contribution ($f_{\text{d}} = 0.0$, top row), the inferred metallicities remain predominantly sub-solar. Incorporating a realistic diffuse IGM component ($f_{\text{d}} = 0.6$, middle row) yields near-solar metallicities across most sightlines, with a few localized values exceeding solar abundance. The model with $f_{\text{d}} = 0.8$ (bottom row), drives the inferred metallicities into the strongly super-solar regime ($Z \gg 1 \, Z_{\odot}$) across several sightlines, which is physically implausible given the expected dilution by IGM accretion and the energetic limits of galactic feedback enrichment mechanisms.
%***Add more on the figure:
%We do not see any systematic trend in the derived CGM metallicities with these parameters.
\newpage
\subsection{{Values and Errors in the CGM metallicity:}}
For an individual sightline $i$, the lower and upper statistical uncertainties are propagated in quadrature using fractional uncertainties scaled by the individual median $Z_{\text MW,i}$:
\begin{align}
\sigma_{i} &= Z_{\text{MW,CGM,i}} \times \sqrt{ \left( \frac{\sigma_{\text{FRB}, i}}{\text{DM}_{\text{FRB}, i}} \right)^2 + \left( \frac{\sigma_{\text{metal}, i}}{\text{DM}_{\text{metal}, i}} \right)^2 } 
\end{align}

Across the ensemble of sightlines $N$ under a given diffuse cosmic fraction model $f_d$:
\begin{enumerate}
    \item \textbf{Systematic Uncertainty:} Defined using the sample median $Z_{{P_{50}}}$ with asymmetric systematic boundaries determined via percentiles:
    \begin{equation}
    Z_{\text{MW}} = Z_{P_{50}} \,_{-\left(Z_{P_{50}} - Z_{P_{16}}\right)}^{+\left(Z_{P_{84}} - Z_{P_{50}}\right)}
    \end{equation}
    \item \textbf{Statistical Uncertainty:} Aggregated across all $N$ sightlines via calculating the root-mean-square:
    \begin{equation}
    \sigma_{\text{stat}} = \sqrt{\frac{1}{N^2} \sum_{i=1}^N \sigma_{i}^2}
    \end{equation}
\end{enumerate}

%***Do you really need to define Z_global? It kind of implies universal, which is not what we are measuring. Can we not call it Z_MW? For previous Z_MW, make it Z_MW,i to designate that to an individual sightline. ***

Across all evaluated models, the statistical uncertainties and the global systematic uncertainty increase substantially as a function of the assumed diffuse fraction $f_d$:

\begin{itemize}
    \item For $f_d = 0.0$, we obtain a global median ratio of $Z_{\text{MW}} = 0.19_{-0.13}^{+0.21} \, (\text{sys}) \pm {0.05}/{0.05} \, (\text{stat})$.
    \item Increasing the diffuse fraction to $f_d = 0.6$ yields $Z_{\text{MW}} = 0.39_{-0.30}^{+0.53} \, (\text{sys}) \pm {0.16}/{0.18} \, (\text{stat})$, representing a factor of $\sim 2$ increase in the median for $f_d = 0.0$.
    \item At $f_d = 0.8$, the global median ratio reaches $Z_{\text{MW}} = 0.49_{-0.39}^{+1.6} \, (\text{sys}) \pm {0.57}/{0.58} \, (\text{stat})$. However, because the lower statistical uncertainty exceeds the central median value ($Z < \sigma_{\text{stat, low}}$), the data cannot establish a bounded physical detection and is instead reported as an upper limit of $Z_{\text{MW}}$, which is $2.09 \, Z_{\odot}$, taking into account the upper statistical and systematic bound. This reflects the severe skewness and loss of constraining power in high IGM diffuse fraction regimes.
    
\end{itemize}
We evaluated the global sample metallicity while accounting for both uncertainties in another way. We implemented an asymmetric bootstrap resampling procedure. In each of $10^5$ iterations, we first perturbed each sightline's median metallicity by drawing a value from a split-normal distribution using its upper or lower $1\sigma$ uncertainty with equal $50\%$ probability. We then generated a bootstrap sample by randomly drawing 25 sightlines from this perturbed set with replacement and computing the sample median. Finally, the $16^{\text{th}}$, $50^{\text{th}}$, and $84^{\text{th}}$ percentiles of the resulting $10^5$ bootstrap medians were extracted to define the global sample metallicity and its asymmetric confidence intervals.
\begin{align}
Z_{\text{MW}} (f_{\text{d}} = 0)   &= 0.12  _{-0.04}^{+0.09} \, Z_{\odot} \\[4pt]
Z_{\text{MW}} (f_{\text{d}} = 0.6) &= 0.26_{-0.12}^{+0.18} \, Z_{\odot} \\[4pt]
Z_{\text{MW}} (f_{\text{d}} = 0.8) &= 0.39_{-0.20}^{+0.26} \, Z_{\odot}
\end{align}
As expected, increasing $f_{\text{d}}$ attributes a larger portion of the observed dispersion measure to $\text{DM}_{\text{IGM}}$, thereby reducing the residual Milky Way halo contribution ($\text{DM}_{\text{MW, CGM}}$) and systematically elevating the inferred MW CGM metallicity $Z_{\text{MW}}$\footnote{We obtained  tighter uncertainties obtained via the bootstrap approach relative to the previous method. Note that the bootstrap method is not sensitive to the number of iterations and does not cause oversampling or artificially narrow the confidence intervals; it simply eliminates numerical noise in evaluating the true width of the distribution. Consequently, bootstrap measures how precisely the global halo median is constrained, whereas the first method preserves the broad empirical scatter driven by CGM anisotropy and extreme outliers in some sightlines.}.

\section{{Conclusion and Discussion}}
\label{sec:conclusion}
\begin{itemize}
    \item \textbf{High Metallicity at $f_d = 0.8$:}  Inferred halo metallicities significantly exceeding the solar value ($Z \gg Z_{\odot}$) across certain sightlines and high-$f_{\text{d}}$ choices are physically implausible for the MW CGM. While the star-forming Galactic disk is enriched near solar metallicity ($Z \approx Z_{\odot}$), diffuse gas in the extended halo is predominantly composed of accretion from the Intergalactic Medium (IGM, $Z \ll 0.1 \, Z_{\odot}$) along with contributions from galactic fountain outflows and stellar feedback. In situ metal enrichment mechanism models, such as core-collapse supernovae and AGB wind injection, cannot feasibly elevate the diffuse, low-density hot phase of the outer halo to several times solar abundance on global scales. Consequently, sightlines yielding $Z \gtrsim 1 \, Z_{\odot}$ indicate that the residual dispersion measure assigned to the CGM ($\text{DM}_{\text{MW, CGM}}$) has been under-predicted due to over-subtraction of the cosmological component ($\text{DM}_{\text{cosmic}}$) or localized host galaxy contributions ($\text{DM}_{\text{host}}$).

However, this does not require a different diffuse cosmic baryon fraction at low redshift. Rather, it reflects the strongly skewed distribution of $\mathrm{DM_{cosmic}}$ for short sightlines, which intersect few intervening halos. Adopting the \cite{Mac2020} distribution with $\sigma_{\rm DM} = F z^{-1/2}$ and $F = 0.2$--$0.32$, we find that at $z = 0.1$ the median $\mathrm{DM_{cosmic}}$ is only $\approx 0.76$--$0.79$ of the mean, and $\approx 74\%$ of sightlines lie below $\langle \mathrm{DM_{cosmic}} \rangle$. For a typical low-redshift sightline the effective diffuse fraction is therefore $f_{\rm d,eff} \approx 0.6$--$0.7$, consistent with the values that yield physical CGM metallicities in our analysis. Subtracting the mean Macquart relation thus systematically over-corrects individual low-redshift FRBs, and $f_{\rm d} \approx 0.8$ should be regarded as an upper limit on the effective value appropriate for $z < 0.3$ sightlines.
    
    \item \textbf{Sightline Anisotropy:} We observe significant anisotropy in the inferred metallicities across different FRB sightlines. This spatial variation may naturally arise if the two sightlines intersect distinct physical regimes within the CGM. Furthermore, spatial offsets between the FRB and absorption column sightlines can sample different density, temperature, and ionization structures, hence different DM values. Taken together, these sightline-to-sightline fluctuations point toward a highly inhomogeneous, multi-phase, and anisotropic metal distribution throughout the extended MW halo, shaped by complex interactions between feedback outflows and cosmic accretion, as also found by other observational studies \citep[e.g.,][]{Gupta2012,Gupta2023Nat,Bhatta2023}.
    
    \item \textbf{Feedback Retention in the CGM:} Even under our most conservative, lower-bound metallicity estimates ($f_{\text{d}} = 0.0$), the inferred CGM metallicity significantly exceeds the weakly enriched values characteristic of purely intergalactic accretion ($Z_{\text{IGM}} \lesssim 0.1 \, Z_{\odot}$). This baseline enrichment demonstrates that the MW halo is not merely populated by infalling primordial gas, but has undergone substantial metal pollution over cosmic time. Outflows from supernova feedback and active galactic nucleus (AGN) activity eject metal-rich gas from the deep potential well of the Galactic disk into the extended halo. The persistent elevation of $Z_{\text{CGM}}$ across our sample confirms that these feedback-driven outflows are efficiently trapped and retained within the gravitational potential of the dark matter halo, serving as a long-term reservoir for recycled baryonic material rather than escaping completely into the diffuse IGM (See also \citealt{Gupta2023Nat}).
\end{itemize}

\begin{table*}
\centering
\caption{Summary of selected local FRBs showing Galactic coordinates ($b, l$), redshift ($z$), observed Dispersion Measures ($\text{DM}_{\text{obs}}$), and inferred CGM metallicity ratios $Z$ ($Z_\odot$) across three diffuse IGM fraction parameters ($f_{\text{d}} = 0.0, 0.6, 0.8$). All values are rounded to two decimal places.}
\label{tab:local_FRB_list_3fds}
\small
\begin{tabular}{lrrrrccc}
\hline\hline
FRB & $b$ ($^{\circ}$) & $l$ ($^{\circ}$) & $z$ & $\text{DM}_{\text{obs}}$ ($\text{pc/cm}^{3}$) & $Z \, (f_{\text{d}}=0.0)$ & $Z \, (f_{\text{d}}=0.6)$ & $Z \, (f_{\text{d}}=0.8)$ \\
\hline
20181220A & -11.49 & 106.82 & 0.0960  & 209.40 & $0.27_{-0.17}^{+0.14}$ & $0.91_{-0.62}^{+0.70}$ & $2.10_{-1.48}^{+1.92}$ \\
20181223C & 79.40  & 207.90 & 0.0850  & 112.50 & $0.43_{-0.30}^{+0.28}$ & $1.42_{-1.07}^{+1.24}$ & $3.26_{-2.51}^{+3.31}$ \\
20190418A & -22.89 & 179.30 & 0.1200  & 184.50 & $0.12_{-0.09}^{+0.10}$ & $0.43_{-0.33}^{+0.44}$ & $1.10_{-0.87}^{+1.28}$ \\
20190425A & 33.08  & 42.10  & 0.0810  & 128.20 & $0.48_{-0.31}^{+0.40}$ & $1.85_{-1.29}^{+1.92}$ & $5.53_{-3.98}^{+6.64}$ \\
20181030A & 40.06  & 134.81 & 0.0500  & 103.00 & $0.39_{-0.28}^{+0.28}$ & $0.91_{-0.68}^{+0.80}$ & $1.44_{-1.09}^{+1.37}$ \\
20200120E & 41.22  & 142.19 & 0.00013 & 87.82  & $0.55_{-0.41}^{+0.44}$ & $0.55_{-0.41}^{+0.44}$ & $0.55_{-0.41}^{+0.44}$ \\
20230203A & 52.88  & 182.20 & 0.1464  & 420.10 & $0.06_{-0.03}^{+0.02}$ & $0.09_{-0.05}^{+0.04}$ & $0.10_{-0.05}^{+0.04}$ \\
20230222A & 12.01  & 207.97 & 0.1223  & 706.10 & $0.39_{-0.38}^{+0.06}$ & $0.46_{-0.45}^{+0.08}$ & $0.49_{-0.48}^{+0.09}$ \\
20230222B & 47.93  & 45.45  & 0.1100  & 187.80 & $0.19_{-0.11}^{+0.13}$ & $0.42_{-0.27}^{+0.33}$ & $0.63_{-0.42}^{+0.55}$ \\
20230311A & 18.06  & 157.94 & 0.1918  & 364.30 & $0.07_{-0.04}^{+0.03}$ & $0.14_{-0.09}^{+0.06}$ & $0.20_{-0.13}^{+0.10}$ \\
20230703A & 67.87  & 137.91 & 0.1184  & 291.30 & $0.05_{-0.03}^{+0.02}$ & $0.08_{-0.05}^{+0.03}$ & $0.09_{-0.06}^{+0.04}$ \\
20230730A & -16.89 & 159.27 & 0.2115  & 312.50 & $0.29_{-0.28}^{+0.08}$ & $<1.39$ & $<3.59$ \\
20230926A & 21.84  & 68.32  & 0.0553  & 222.80 & $0.31_{-0.21}^{+0.16}$ & $0.43_{-0.30}^{+0.25}$ & $0.50_{-0.35}^{+0.29}$ \\
20231005A & 44.52  & 50.15  & 0.0713  & 189.40 & $0.19_{-0.11}^{+0.14}$ & $0.30_{-0.19}^{+0.23}$ & $0.37_{-0.23}^{+0.29}$ \\
20231011A & -20.06 & 126.79 & 0.0783  & 186.30 & $0.19_{-0.17}^{+0.15}$ & $0.38_{-0.36}^{+0.35}$ & $0.55_{-0.51}^{+0.52}$ \\
20231017A & -20.21 & 101.99 & 0.2450  & 344.20 & $0.05_{-0.03}^{+0.02}$ & $0.16_{-0.11}^{+0.08}$ & $0.34_{-0.25}^{+0.22}$ \\
20231025B & 27.67  & 93.30  & 0.3238  & 368.70 & $0.08_{-0.04}^{+0.03}$ & $0.32_{-0.21}^{+0.18}$ & $1.32_{-0.98}^{+1.10}$ \\
20231123A & -17.51 & 200.75 & 0.0729  & 302.10 & $0.05_{-0.03}^{+0.04}$ & $0.07_{-0.05}^{+0.06}$ & $0.08_{-0.05}^{+0.07}$ \\
20231128A & 73.18  & 118.00 & 0.1079  & 331.60 & $0.04_{-0.02}^{+0.01}$ & $0.06_{-0.03}^{+0.02}$ & $0.06_{-0.04}^{+0.03}$ \\
20231201A & -22.37 & 163.63 & 0.1119  & 169.40 & $<1.27$ & $<6.30$ & $<23.95$ \\
20231204A & 66.86  & 101.44 & 0.0644  & 221.00 & $0.08_{-0.05}^{+0.04}$ & $0.11_{-0.07}^{+0.06}$ & $0.12_{-0.08}^{+0.07}$ \\
20231206A & 27.53  & 160.03 & 0.0659  & 457.70 & $0.06_{-0.03}^{+0.02}$ & $0.07_{-0.03}^{+0.03}$ & $0.07_{-0.04}^{+0.03}$ \\
20231223C & 33.72  & 49.33  & 0.1059  & 165.80 & $0.27_{-0.17}^{+0.20}$ & $0.83_{-0.56}^{+0.76}$ & $1.70_{-1.19}^{+1.73}$ \\
20231229A & -26.11 & 133.52 & 0.0190  & 198.50 & $0.15_{-0.13}^{+0.11}$ & $0.17_{-0.15}^{+0.13}$ & $0.17_{-0.15}^{+0.14}$ \\
20231230A & -22.75 & 198.54 & 0.0298  & 131.40 & $0.26_{-0.20}^{+0.25}$ & $0.39_{-0.30}^{+0.40}$ & $0.45_{-0.35}^{+0.47}$ \\
\hline
\end{tabular}
\end{table*}

\normalsize
\section*{Data Availability}
{We use Python packages of astropy \citep{Astropy2022}, numpy \citep{numpy}, pyGEDM \citep{pyGEDM}, for this work.}
\section*{acknowledgements}
M.R. acknowledges support from the Center for Cosmology and Particle Physics (CCAPP) fellowship at Ohio State University. M.R. thanks Todd Thompson for the useful discussion. S.D. acknowledges support provided by NASA through Hubble Fellowship grant HST-HF2-51551.001-A, awarded by the Space Telescope Science Institute, which is operated by the Association of Universities for Research in Astronomy, Inc., for NASA, under the contract NAS 5-26555.
S.M. is grateful for the NASA ADAP grant 80NSSC22K1121.

\label{lastpage}
%\bibliography{bridging_scales}{}
\bibliography{main}
\bibliographystyle{aasjournal}
\end{CJK}
\end{document}